# Emergence of the protein universe from organismal evolution

Konstantin B. Zeldovich[1], Boris E. Shakhnovich[2], Eugene I. Shakhnovich[1]

[1]Department of Chemistry and Chemical Biology, and [2]Molecular and Cellular Biology
Harvard University, 12 Oxford Street, Cambridge, MA 02138

Abstract

In this work we propose a physical model of organismal evolution, where phenotype - organism life expectancy - is directly related to genotype – the stability of its proteins which can be determined exactly in the model. Simulating the model on a computer, we consistently observe the ''Big Bang'' scenario whereby exponential population growth ensues as favorable sequence-structure combinations (precursors of stable proteins) are discovered. After that, random diversity of the structural space abruptly collapses into a small set of preferred structural motifs. We observe that protein folds remain stable and abundant in the population at time scales much greater than mutation or organism lifetime, and the distribution of the lifetimes of dominant folds in a population approximately follows a power law. The separation of evolutionary time scales between discovery of new folds and generation of new sequences gives rise to emergence of protein families and superfamilies whose sizes are power-law distributed, closely matching the same distributions for real proteins. The network of structural similarities of the universe of evolved proteins has the same scale-free like character as the actual protein domain universe graph (PDUG). Further, the model predicts that ancient protein domains represent a highly connected and clustered subset of all protein domains, in complete agreement with reality. Together, these results provide a microscopic first principles picture of how protein structures and gene families evolved in the course of evolution.

**Introduction**

Molecular biology has collected a wealth of quantitative data on protein sequences and structures, revealing complex patterns of the protein universe, such as a scale-free behavior of structural similarity, and the markedly uneven usage of protein folds [1; 2; 3; 4; 5]. On a much higher level of biological hierarchy, ecology, evolution theory, and population genetics established a framework for studying speciation, population dynamics and other large-scale biological phenomena [6; 7; 8]. While it is widely accepted that gene families and the Protein Universe emerged during the course of molecular evolution through selection [9; 10; 11], there is a substantial gap in our conceptual and mechanistic understanding of how molecular evolution occured or what are the determinants of selection. Indeed, evolution, as we understand it, proceeds at the level of organisms and populations but not at the level of genomes. Evolutionary selection at the molecular level occurs due to a relation between genotype and phenotype, although a detailed understanding of this relation remains elusive.

A number of phenomenological models (e.g. quasispecies) were developed where fitness of an organism was defined using the sequence of a genome. [12; 13; 14]. However, the relationship between genotype and phenotype in quasispecies and similar population genetics models is purely phenomenological. For example, in single-fitness peak models, one specific genotype is postulated to be most fit while deviations from it confer selective disadvantage. Despite providing several important insights, this kind of approach lacks a fundamental microscopic connection between fitness and easily justifiable and measurable quantities (e.g. structure/stability, function or regulation) of proteins. Therefore, these models cannot accurately describe molecular evolution of proteins.

On the other hand a number of models were proposed that focus on emergence and evolution of protein folds under direct pressure on their molecular properties such as stability [11; 15; 16; 17; 18], folding kinetics [19; 20] and mutational robustness [21]. Schuster and Stadler [22] first studied evolution of macromolecules – RNA in the context of population dynamics. In a series of papers [10; 11] Taverna and Goldstein used Eigen model of reaction flow to grow populations of proteins modeled as two-dimensional 25-mers having two types of amino acid residues. These authors showed that when requirement to exceed certain stability threshold is imposed the resulting distribution of structures in the evolved population appears highly skewed towards more designable [23] structures.

One of the most surprising features of the Protein Universe, is an uneven and broad distribution of proteins over folds, families and superfamilies. While this fact had been noted by many researchers long ago [1; 3; 24; 25], the quantitative descriptions of such uneven distributions began to emerge only recently. Huynen and Nimwegen reported that sizes of paralogous gene families follow power-law distribution [5]. Gerstein and coworkers [4] observed power-law distribution of frequencies of several other properties of gene families as defined in the SCOP database [26]. Dokholyan et al [2] studied a network of structural similarities between protein domains (called Protein Domain Universe Graph, or PDUG) and found that distribution of connectivities within such graph follows power law as well (within a limited range of connectivity variance) making the PDUG a finite size counterpart of a scale-free network

Ubiquitous nature of power law dependencies of many characteristics of Protein Universe may suggest its possible common origin from fundamental evolutionary dynamics and/or physics of proteins. Huynen and Nimwenger [5], Gerstein and coworkers [4] and Koonin and coworkers [9; 27; 28] proposed dynamic models (the version proposed in[27] is called BDIM) based on gene duplication as a main mechanism of creation of novel types. Such models, while providing power-law distribution of family sizes in some asymptotic cases, are based on several assumptions that call into question their generality. In particular, as pointed our by Koonin and coworkers, in order for gene duplication dynamic models to provide non-trivial power law distributions of paralogous family sizes, one has to assume that probability of gene duplication *per gene* depends, in a certain regular way, on the size of already existing gene family. Further, even under this assumption power-law distribution in the BDIM model arises only asymptotically in a steady state of evolutionary dynamics [28]. In contrast, the duplication and divergence phenomenological model of Dokholyan et al [2] did not use such dramatic assumptions. However, this model is limited to explanation of scale-free nature of PDUG and does not provide any insight as to nature of power-law distribution of gene family sizes. In any case, models like the ones proposed in [2; 4; 5; 27] and other works are purely phenomenological in nature whereby proteins are presented as abstract nodes and where sequence-structure relationships are not considered.

An alternative explanation of the uneven distributions observed in Protein Universe posits that certain intrinsic properties of protein structures may confer selective advantage to the organism. In particular, it has been noted that the number of sequences that can fold into a protein structure, or structure's designability [23; 29; 30; 31], varies greatly between structures. However a careful analysis of the possible distribution of fold family sizes due to variation in protein designability predicts exponential distribution in family sizes [32], not power law observed in Nature. Furthermore, variation in designability alone cannot explain the ''scale-free'' nature of PDUG [32].

Here, we present a microscopic model of organismal evolution (Figure 1) with realistic generic population dynamics scenario where fitness of an organism is related to the ability of its proteins to be in their native conformations. Since the latter can be estimated exactly in our model from sequences of evolving genomes, the proposed model provides a rigorous, microscopic connection between molecular evolution and population dynamics. We demonstrate that the model indeed bridges multiple time scales, thus providing an insight on how selection of a best-fit phenotype results in molecular selection of proteins and formation of stable, long-lasting protein folds and superfamilies. The resulting Protein Universe matches quantitatively the real one including such key properties as power-law distribution for gene family and superfamily sizes and ''scale-free''-like PDUG. The proposed model can be viewed as a first step towards a microscopic, first principles description of evolution of Protein Universe.

**Results and Discussion.**
**Population dynamics, fold discovery, and punctuated evolution.**

Our evolution dynamics runs start from initial population of 100 organisms each having the same one primordial gene in their genomes. Initial gene sequence is random. Runs proceed according to evolutionary dynamics rules as described in Model and Methods section (see also Figure 1). The life expectancy of an organism is directly related to stability of its proteins as explained in Methods section. This is equivalent to a postulate that all genes of primordial organisms are essential.

We found that out of 50 simulation runs starting with different starting sequences, 27 runs successfully resulted in a steady exponential growth of the population, whereas in 23 runs the population has quickly gone extinct. A typical behavior of the population growth and protein structure dynamics in a successful evolution run is shown in Figure 2. After a period of "hesitation" lasting for about 100 time steps, a steady exponential growth of the population sets in (Figure 2b). The characteristic sawtooth time dependence of population size is due to the decimation procedure to keep the population within computationally accessible limits. In Figure 2c, we present the mean native state probability $<P_{nat}>$ of all proteins present in the population at a given time. Due to mutations and selection, $<P_{nat}>$ steadily increases with time, and dramatically exceeds the mean $P_{nat}$ for random sequences, $<P_{nat}^{rand}>=0.23$. In contrast to earlier models [11] the selection pressure is applied to whole organisms rather than to individual protein molecules. The genotype-phenotype feedback, which we model by eq. (2) (see Methods), transfers the pressure from organisms to individual proteins to gene sequences. Figures 2(b,c) show that our selection mechanism works and results in the discovery of stable proteins due to evolutionary pressure.

Using our model, we can follow each structure in the population. In Figure 2a color hue encodes the number of genes in the population corresponding to each of the 103346 lattice structures (ordinate) as a function of time (abscissa). Structures marked in red are the most abundant in population at a given time, while cyan background corresponds to structures not found in any of the evolving organisms. The most important feature of this plot is the appearance of specific structures that correspond to highly abundant proteins comprising a significant fraction of the gene repertoire of the population. In what follows we will call them Dominant Protein Structures (DPS). Such proteins visually appear as bright lines on Figure 2a. What is the genesis of DPS and how is their appearance related to population growth or decay?

To answer this, let us track the development of the population of structures in time by comparing the structure repertoire, the population size and $<P_{nat}>$ plots. At $t=0$ the proteome consists of a single sequence-structure combination (a single line on the structural repertoire plot) which corresponds to all individuals in the initial population having that structure in the genome. Over time, random mutations diverge sequences in each organism such that the dominance of a single structure is lost. This can be seen as a smeared line on the structural repertoire plot, as shown in Figure 2a, $t$<100. However, at a certain point (vertical dashed line in Figure 2), very favorable sequence-structure combinations are discovered. They represent DPS whose incorporation into the genome leads to an abrupt increase of <$P_{nat}$> and explosive exponential growth of the population through increase in fitness. Shortly after the discovery of that DPS, the diversity of the structural space abruptly collapsed, as most of the organisms converge towards the newly discovered DPS. Therefore, as we observe in this model, the "Big Bang" in population dynamics is directly related to the discovery of specific protein folds. As seen on Figure 2a, these folds are very persistent in time. Nevertheless, fold discovery can occur at later stages of evolution. For example, in this particular simulation, at $t$~800, new folds were discovered (short dashed line in Figure 2a), they become new DPS and the initial DPS are completely replaced by the new folds by $t$~1000. This switchover, accompanied by a marked increase of <$P_{nat}$>, is a clear manifestation of punctuated discoveries of new folds coupled with selection at the organismal level.

Even though the number of organisms increases exponentially, the number of genes in each genome increases very slowly (and stabilizes after the discovery of DPS (Supplementary Figure 1, red curve). Indeed, large genomes are not very advantageous in our model, as mutations occur in all of the genes whereas the death rate is controlled by the gene with the lowest $P_{nat}$. Thus, it is only this gene that bears the brunt of selective pressure. Therefore, the rest of the genome accumulates mutations and is more prone to deleterious mutations. Unless all of the genes are very carefully selected (or formation of pseudogenes is allowed), larger number of genes means that there is a substantial probability that a point mutation will result in a sequence-structure combination with a very low $P_{nat}$, immediately killing the organism. The observed slow increase of the size of the genome reflects the subtle balance between the selection pressure and gene duplication.

Supplementary Figure 2 shows the structural repertoire and population size of an unsuccessful simulation run, where the population quickly became extinct. This simulation did not result in discovery of a stable fold, and the structural space was evenly filled till the extinction of the population. We found (data not shown) that the choice of starting sequence does not have any significance in determining whether a particular simulation run will result in exponential growth or extinction. Furthermore, in the case of most unsuccessful evolution runs, the genome size rapidly increases with time (Supplementary Figure 1, blue curve), decreasing the average evolutionary pressure per gene and further complicating the discovery of DPS.

Based on these observations, we conjecture that biological evolution, exponential population growth, and existence of stable genomes are possible only after the discovery of a narrow set of specific protein structures. Unfortunately, at present we do not know what properties of a structure make it a potential DPS and what is the role of the particular sequence of events that resulted in the emergence of DPS. Designability [23] or maximum eigenvalue of the structure's contact matrix [30; 32] do not seem to be major factors in determining whether a structure is a potential DPS.

**Emergence of Families and Superfamilies**

To quantify the persistence of the DPS during evolution, we calculated the distribution of DPS lifetimes, defined as the time span during which a structure comprises more than 20% of the genes present in the population– i.e. time between emergence of a DPS and its extinction in the population (see Fig.3a). We consider only DPSs that already completed their ''lifecycle'' i.e. a DPS that emerged and went extinct over the time of an evolutionary simulation. It is clear from Fig.3b that the life-time of the DPS is much greater than that of an organism, or the average time between successive mutations. Moreover, the distribution of DPS lifetimes clearly follows a power-law-like distribution. The long non-exponential tail of the distribution demonstrates that some protein folds are extremely resistant to mutations and may persist over thousands of generations. Over such a long time, diverse protein (super)families are formed around the DPS folds. This is illustrated on Fig.4a that shows distribution of sizes of evolved families and superfamilies of proteins. To avoid confusion we note that families and superfamilies here are defined not necessarily as sets of paralogous sequences but in the same way as they are defined in SCOP [26] : protein families are defined as sets of all (not necessarily belonging to the same organism) homologous sequences that fold into a given domain structure and superfamilies are defined as all monophyletic sets of sequences whose homology may not be detectable by sequence comparison methods but which nevertheless fold into structurally similar domains. The statistics of protein families is dominated by orthologous genes, in contrast to paralogous families studied in [4; 9]. As can be seen in Fig.4a both family and superfamily size distributions of evolved proteins follow almost perfect power laws with power law exponent being greater for superfamilies (-2.92) than that for families (-1.77).

In order to compare this result with real proteins we plot here the distribution of family and superfamily sizes of real proteins. As a measure of family sizes here we estimated the number of homologous sequences that fold into a given domain (see Methods) and as a proxy for superfamily size we estimated the number of functions performed by each domain. Clearly the distributions in Fig.4b follow power-law statistics, and as in model, the exponent for the superfamily distribution (-2.2) is greater than that for families (-1.6). Quantitatively, the slopes of the model and real distributions match each other quite well.

**Structural similarity network of evolved proteins.**

An important characteristic of the set of evolved proteins is the protein domain universe graph (PDUG) [2]. In this graph, non-homologous proteins are clustered according to the degree of their structural similarity, which should exceed a certain threshold. It is known [2] that in natural proteins, the size of the largest cluster (giant component) of the PDUG abruptly shrinks at some value of the threshold, similar to the percolation transition. The degree distribution of the graph, i.e. the probability *p(k)* that a protein has *k* structurally similar neighbors, is a power law at the transition point. The scale-free character of this graph is believed to be a consequence of divergent evolution [2; 33; 34] as suggested by simple phenomenological "duplication and divergence" models [2]. Therefore, it is important to test whether our model can reproduce the global features of the natural protein universe that are manifest in the unusual properties of the PDUG.

Here we plot the PDUG of evolved proteins using Q-score – the number of common contacts between a pair of proteins – as a structural similarity measure [34]. The degree distribution of the evolved PDUG at similarity threshold $Q$=17 (the mid-transition in giant component of the evolved graph, see Supplementary Figure 3) is shown in Figure 5a. The degree distribution plot clearly shows that the graph consists of two components, a scale-free-like component at lower *k*, and a small but very highly connected component at high $k$. As a control, we computed $p(k)$ for a divergent model without the genotype-phenotype feedback, with the fixed death rate of organisms equal to the death rate in the exponential growth regime of evolution model. The degree distribution of the PDUG obtained in this control simulation is shown in Figure 5b. The control graph is weakly connected, indicating randomness of the discovered structures. The degree distribution of the control graph is well approximated by a Gaussian distribution, in contrast to the one obtained from simulation or the empirical one computed from available data.

Therefore, evolutionary selection has a profound effect on the global structure of evolved protein universe. In the model, the structural similarity graph (PDUG) splits into a scale-free-like part and a highly connected part, corresponding to the DPS, populated by many dissimilar sequences. To further characterize this graph, we plot the clustering coefficient $C(k)$ of the node as function of its degree $k$ (Fig 6a). The overall trend is that the highly connected nodes are also highly clustered, i.e. their neighbors tend to be connected to each other. Notably, the DPS provide extreme values of both connectivity and clustering coefficient, much higher than those of the bulk of structures. A very similar picture is obtained in natural PDUG, Figure 6b, where clustering coefficient $C(k)$ is also positively correlated with node degree $k$. We note that positive correlation of clustering coefficient with node degree is a unique property of PDUG that stems from its evolutionary dynamics. In other networks, e.g. protein-protein interaction nets, the clustering coefficient is negatively correlated with connectivity [35]. In other words, the peculiar property of the network of the protein structure similarities is that it is highly clustered at any value of connectivity $k$. The early "saturation" of $C(k)$ with increasing $k$ clearly illustrates this unusual feature. How could this feature of PDUG emerge in divergent protein evolution?

In Figure 7, we illustrate the divergent evolution scenario as realized in our model. Divergence and selection lead to the infrequent discovery of new protein folds (dashed circles). Within these folds, mutations result in the formation of protein (super)families. The size of protein families steadily increases with time, so older families are generally larger. However, fold formation can occur at any time, branching off any family, so the newly formed families will be necessarily small. At the same time, the structures corresponding to superfamilies are all pairwise similar to each other and for that reason they are highly clustered in the PDUG. Therefore, at any moment, the snapshot of the evolving protein universe will comprise tightly clustered families of all sizes. This is the scenario of divergent evolution found in our model, which is compatible with the experimental observations (Figure 6).

Our simulations predict that new folds emerge as offsprings of DPS. In this picture DPS serve as prototypes of first ancient folds. As seen in Fig.6a, folds representing DPS are highly connected and highly clustered ones. Following this logic one should expect that ancient protein folds, being closer to prototypical DPS should be highly clustered and more connected than later diverged folds. To test this prediction we analyzed the subgraph of PDUG corresponding to last universal common ancestor (LUCA) domains [36]. There are 915 LUCA domains. We compared the connectivity and clustering coefficient in the PDUG subgraph corresponding to LUCA domains with distributions for the same characteristics for 915 randomly selected domains as a control. The null hypothesis is that a random subset of protein domains has connectivity and clustering coefficients similar to that of the LUCA domains. In Figure 8 we present the histograms of mean connectivity and clustering coefficient found in 20000 subsets of N=915 randomly chosen protein domains (out of total of 3300 DALI domains constituting the PDUG, see [2]) . For random subsets of 915 domains from the PDUG, $<k>$=2.91, $<C>$=0.197 while the average values of the same parameters for the 915 LUCA protein domains: $<k>$=4.61, $<C>$=0.267. The values of $<k>$ and $<C>$ for the LUCA domains are marked by red lines in Figure 8, and they are far to the right from the maxima of distributions of $<k>$ and $<C>$ for randomly chosen domains, yielding extremely low $p$-values ($p<10^{-10}$) that LUCA domains are connected and clustered just as a random subset of the PDUG (assuming Gaussian distributions of mean connectivities and clustering coefficients for random subsets of the PDUG Fig.8). This proves that LUCA domains are statistically more connected and clustered than an equivalent set of random protein domains as predicted from our simulations.

**Conclusions.**

In this work we introduced a model of divergent evolution which directly relates evolving protein sequences and structures to the life expectancy of the organism. We have used a simple physical model of protein thermodynamics, and a simple model of the population dynamics. The main assumption of our minimalistic model is that the necessary condition of survival of a living organism is that its proteins adopt their native conformations. Therefore, death rate of the organisms decreases when their proteins become more stable against thermal denaturation or unfolding. In other words we assume that *all genes of our model organism are essential*.

There is a common belief that the experimentally observed moderate stability of natural proteins is a result of positive selection on function. As a ''proof'' of this conjecture a circular argument is offered that natural proteins are not extremely stable. However, there is no experimental or logically flawless proof of this conjecture. Just the opposite - recent study demonstrated that higher stability of a protein confers a selective advantage to a protein by making it more evolvable, by enhancing its ability to tolerate more mutations and as a result evolve a new function [37]. A more plausible explanation of moderate stability of natural proteins is that it is a direct result of a tradeoff between stability in the native conformation and entropy in sequence space that opposes an evolutionary optimization beyond necessary levels [38]. We observe exactly this phenomenon in our model: while organisms with more stable proteins have selective advantage, the opposing factor – enormity of search in sequence/structure space – results in a compromise level of stability which corresponds to stable but not overstabilized proteins (see Figure 2c).

Unlike in many previous attempts, our model explicitly describes the interplay of evolution of individual genes and that of genomes (organisms) as a whole, since death of an organism leads to a complete loss of its genome. The model gives important insights into the interplay between molecular evolution, protein fold evolution, and population dynamics. In combination with selection pressure, random diffusion in sequence and structure spaces eventually leads to the discovery of specific structures, DPS, that are resistant to mutations and form very evolvable proteins. This, in turn, immediately leads to the "Big Bang" of exponential population growth, as mutations are no longer a big threat to viability. The DPS persist over many generations, and may be infrequently replaced or augmented by other, even more favorable, structures, in a process similar to punctuated evolution. The remarkable separation of timescales between frequent mutations and rare DPS formation allows for the formation of the protein universe, superfamilies, and families.

The model and simulations presented here provide a quantitative first-principles description of evolution of Protein Universe. Despite simplicity of the structural model of proteins and phenotype-genotype relation invoked, it is able to reproduce quantitatively all power-law distributions that are observed in natural Protein Universe – in sizes of orthologous families and superfamilies and scale-free like distribution of connectivities in PDUG. This is the most striking key result of this work. Earlier phenomenological models succeeded in describing some aspects of power-law behavior, but not all of them in one model and always at the expense of invoking dramatic assumptions about dependence of rates of gene duplication on sizes of already existing gene families. Here no such assumptions are made as the model is fully microscopic in nature. The most intriguing (and relevant) question is what is the origin of the universally observed power-law distributions in this model? Clearly an explanation proposed in many phenomenological models [27] [5] is not applicable here because the rate of all processes, including gene duplication is constant in the model and does not depend on sizes of already existing gene families. Therefore there is nothing peculiar in the gene birth/death dynamics in the model that could result in power-law distributions. An only plausible reason may be that the underlying dynamics in sequence and structure spaces, coupled with selective pressure, is responsible for the emerging power law distributions. Indeed, our key finding concerns *dynamics* of fold discovery and death - that the lifetimes of DPS are power-law distributed (Fig.3b). The size of a protein family (and superfamily, on longer time scales) is proportional to DPS lifetime as illustrated on Fig.7. Indeed, power law exponents for family size distribution and DPS lifetimes are very similar. It was noted that sequence space statistics is equivalent to statistics of a complex spin model [39,40]. Correspondingly dynamics in sequence and structure spaces may, under strong selective pressure, – which is equivalent to low temperature in a spin model – exhibit glassy behavior that is characterized by a broad distribution of relaxation times giving rise to power-law distribution of DPS lifetime. While these observations are suggestive, a more detailed future analysis will make it possible to find a definite answer as to the origin of ubiquitous power law distributions in sequence and fold statistics.

Our model of natural selection is minimalistic and is limited in its scope. It does not take into account such important biological processes as gene recombination, sexual reproduction, and Darwinian selection due to competition of populations for limited resources. However, we believe that it is an important step towards the unification of microscopic physics-based models of protein structure and function and the macroscopic (so far, phenomenological) description of the evolutionary pressure. Its extensions are straightforward and may include a more explicit consideration of protein function, protein-protein interactions and fitness function that rewards functional (and therefore, structural) innovations. Furthermore, since habitat temperature enters the model explicitly it can be used to study thermal adaptation of organisms. This work is in progress.

**Model and Methods.**

**Population dynamics and genotype-phenotype relationships**. We assume that for an (early) organism to function properly, it is imperative that its proteins spend significant part of the time in their native conformations at a given environmental temperature. Let an organism be represented by its genome, and let $P_{nat}^{(i)}$ be the thermal probability that protein $i$ is in its native conformation. As a simplest approximation, we assume that the probability that an organism is alive is proportional to the lowest $P_{nat}^{(i)}$ across all of its proteins:

$$P_{alive} \propto \min_i P_{nat}^{(i)} , \qquad (1)$$

i.e. longevity of an organism is determined by the least stable protein in the genome ("weakest link" model).

Our model of population and genome dynamics includes four elementary events: 1) random mutation of a nucleotide in a randomly selected gene, with constant rate $m$ per unit time per DNA length; mutations leading to the stop codon are rejected to ensure the constant length of protein sequences; 2) duplication of a randomly selected gene within an organism's genome, with constant rate $u$; 3) birth of an organism via duplication of an already existing organism with constant rate $b$ (the genome is copied exactly); 4) death of an organism, with the rate $d$ per unit time (Figure 1).

In these terms, condition (1) translates into the dependence of organism death rate $d$ on the stability of its proteins:

$$d = d_0 \left(1 - \min_i P_{nat}^{(i)}\right), \qquad (2)$$

where $d_0$ is the reference death rate. This relation gives rise to an effective selection pressure on proteins since organisms which have at least one unstable protein live shorter and thus produce less progeny. This simple, direct and physically plausible relationship between the genotype (thermodynamic properties of the proteins) and the phenotype (life expectancy) is the key novel feature of our model. Another implication of this relationship is that genes do not evolve independently: a very unfavorable mutation in a gene will likely lead to a quick death of an organism, so its complete genome will not be able to proliferate. Such cooperativity creates an important selection pressure towards mutation-resistant genes encoding stable and evolvable (see below) proteins. Interestingly, purely physical factors ensure that resistance to mutations, evolvability of a new function and thermostability are well correlated [37; 38], so little or no trade-off may be needed to satisfy both requirements. To ensure that a sufficient selection pressure is applied, we set $d_0=b/(1-P_{nat}^{(0)})$, where $P_{nat}^{(0)}$ is the native state probability of a protein encoded by a "primordial" gene, which is a single gene in all organisms from which evolution runs start. Therefore, the Malthus parameter $b-d$ of population growth is zero for neutral mutations (not changing $P_{nat}$ with respect to the primordial sequence), positive for favorable mutations which increase $P_{nat}$, and negative for deleterious mutations. In principle, the relationship between growth rate and protein stability can be experimentally verified by analyzing the growth rate of bacteria at elevated temperatures. While the exact biochemical mechanisms leading to slower replication and eventual death are complicated, they all originate in the loss of protein function or enzymatic activity due to thermal denaturation [41]. A conceptually similar sequence evolution model, also using the protein stabililty $P_{nat}$ as fitness parameter has been recently proposed by Goldstein and coworkers [42].

**Simulation algorithm**

In our model, each organism is represented by a list of its genes, 81-nucleotide sequences that are translated into amino acid sequences according to the genetic code. There can be up to 100 genes per organism; the gene duplication rate is chosen so that this limit is never reached in a simulation; typically, organisms have less than 10 genes each at the end of a simulation. Initially, 100 organisms are seeded with one and the same primordial gene; $P_{nat}^{(0)}$ is the native state probability of the protein encoded by the primordial gene.

At each time step of the evolution, each organism can undergo one of the five events: no event at all, or the four events described in the main text (duplication of an organism with probability $b=0.15$, death with rate $d$, gene duplication with probability $u=0.03$, point mutation of a randomly chosen gene with probability $m=0.3$ per gene). The organism death rate is calculated according to eq. (2) in the main text, $d = d_0 \left(1 - \min_i P_{nat}^{(i)}\right)$, with $d_0 = b/(1-P_{nat}^{(0)})$.

Every 25 time steps, an entire set of genes of all currently living organisms is recorded for analysis. The simulation stops after 3000 time steps. Whenever the population size exceeds 20000, we randomly select 2000 organisms for further development and discard the remaining 18000.

**Protein Model.** To simulate the thermodynamic behavior of evolving proteins, we use the standard lattice model of proteins which are compact 27-unit polymers on a 3x3x3 lattice [43]. The residues interact with each other via the Miyazawa-Jernigan pairwise contact potential [44]. It is possible to calculate the energy of a sequence in each of the 103346 compact conformations allowed by the 3x3x3 lattice, and the Boltzmann probability of being in the lowest energy - native - conformation,

$$P_{nat}(T) = \frac{e^{-E_0/T}}{\sum_{i=0}^{103345} e^{-E_i/T}}, \qquad (3)$$

where $E_0$ is the lowest energy among the 103346 conformations, and $T$ is the environmental temperature (in the simulation, we assumed $T=0.5$ in Miyazawa-Jernigan dimensionless energy units).

Simulations were started from 100 identical organisms, each containing one and the same primordial gene, and evolving in time according to the population dynamics rules described above. As population dynamics of our model includes an exponential growth regime, the number of organisms in the simulation must be somehow controlled in order to be able to run evolutionary simulations for a significantly long time. Each time the number of organisms exceeded 20000, populations were decimated by randomly selecting 2000 organisms for further development and discarding the remaining 18000. We explicitly checked that the decimation procedure does not introduce noticeable genetic drift or other artifacts, and the 2000 remaining organisms do carry a representative set of genes on to the future generations.

**Protein domain universe graph**

To construct the protein domain universe graph (PDUG) from the simulation data, we consider only the nonhomologous amino acid sequences. The selection is based on the Hamming distance between the sequences, which should exceed 18 (i.e., less than 33% sequence identity).

To calculate the structure similarity in the PDUG, we use the Q-score similarity measure. The Q-score measure between the two structures $i$ and $j$ is the number of all pairs of monomers $(k,m)$ that are present both in structure $I$ and structure $j$. As there is always 28 contacts in compact 27-mers, Q-score varies from 0 for completely dissimilar structures to 28 for two identical structures. The Q-score is analogous to the DALI Z-score, used as a structural similarity measure for real proteins.

**Family and Superfamily Size Estimate** We take sequences of all structurally characterized domains from HSSP[45]. We use BLAST[46] with threshold 1e$^{-10}$ to identify all sequences with significant homology to each HSSP domain in a non-redundant sequence database NRDB90[47]. We combine each set of sequences with homology into a single gene family. The number of non-redundant sequences matching the domain is the number considered in that family. We then use cross-indexing between NRDB90[47], Swiss-Prot[48] and InterPro[49] to define the set of different functions each gene family performs. The number of different functions as defined by InterPro becomes the number of superfamilies folding into the same domain.

**Definition of LUCA domains** The simplest construction of the LUCA that still yields useful information is the delineation of the very old domains. Any domain shared by the three kingdoms of life can be placed in the last universal common ancestor (LUCA)[50]. If any such domain were not placed in the LUCA, multiple independent discovery (or horizontal transfer) events would be required to explain the occurrence of this domain in all kingdoms. The "extra" evolution involved in this case would result in a less parsimonious scenario. Inclusion of other domains is more probabilistic and depends on the exact form and method of parsimony construction used.[50] We thus define the structural content of the LUCA to be all domains that have homologs in at least one archaeal, at least one prokaryotic and at least one eukaryotic species. This yields approximately a third of the PDUG members.

**Acknowledgments**

The authors acknowledge the stimulating discussions with I.N. Berezovsky and financial support from the NIH.

**Figure captions**

**Figure 1.** Schematic representation of the genome and population dynamics in the model. Individual genes undergo mutations and duplications. Organisms as a whole can replicate, passing their genomes to the progeny, or die, effectively discarding the genome.

**Figure 2. (a).** Structural repertoire of an exponentially growing population as a function of time (abscissa). Ordinate represents the number of the structure out of the 103346 possibles, and abundance of a structure at a given time is encoded by color. Red color corresponds to abundant structures, and cyan to rare or nonexistent ones. Black dotted lines delineate the discoveries of dominant protein structures (DPS, "bright lines" in the structure repertoire). **(b)** Population as a function of time. Exponential growth sets in as soon as stable dominant protein structures have been found. The sawtooth pattern is due to artificial limiting of the exponentially growing population (see text). **(c)** Mean native state probability $<P_{nat}>$ as a function of time.

**Figure 3.** Distribution of life times of DPS. **(a)** Lifetimes are defined as a span between emergence of a DPS when takes over at least 20% of gene population (seen as bright line here) till its extinction as a DPS when it no longer dominates the population. **(b)** The lifetime distribution of DPS approximately follows a power law with exponent -1.87. DPS folds persist over many generations and eventually give rise to protein superfamilies. The blue line indicates the mean life time of an organism.

**Figure 4.** Distribution of family and superfamily sizes (a) model evolution. The blue triangles represent the number of sequences folding into the same structure (gene family); the blue solid line approximates a power law with exponent -1.77. The red circles represent the distribution of the number of nonhomologous (Hamming distance greater than 56%) sequences folding into the same structure (superfamilies). The red solid line is a power law with exponent -2.92. **(b)** Orthologous gene family and superfamily sizes in real proteins. The red circles are the number of different functions performed by each domain as defined by InterPro( Bin size =2, Pearson R= .97 of fit with slope = -2.2) and theblue squares are the number of non-redundant sequences folding into each domain. (Bin size = 10, Pearson R=.92 of fit with slope = -1.5).

**Figure 5** Degree distribution of structure similarity graph (PDUG) in the evolution model (a) and in control (b) where genotype-phenotype relationship does not exist. The similarity threshold was set to $Q=17$ corresponding to the transition point in the largest cluster size (the giant component) of the graph. The slope of the linear approximation of the degree distribution in the evolution model is -1.4 for $\ln k < 4$.

**Figure 6. (a)** Clustering coefficient $C(k)$ vs. node degree $k$ in the structure similarity graph for evolution simulations (black dots) and for control simulations (red circles) without the connection between growth rate and protein stability. **(b)** Clustering coefficient $C(k)$ vs. node degree $k$ for the natural PDUG. Both plots demonstrate a positive correlation between $C(k)$ and $k$.

**Figure 7.** Schematic representation of the formation of protein folds and superfamilies by punctuated jumps in the divergent model. Invention of new folds and their spread in population is a rare event whose time scale exceeds lifetime of an organisms and mutation time scale. On a shorter timescale mutations that do not change protein structure significantly occur and fix in the population. That gives rise to protein families (on the shortest time scales) or superfamilies (on time scales longer than mutational but shorter than fold innovation). Evolutionary time increases from left to right.

**Figure 8**. Probability distribution of the average connectivity (a) and clustering coefficient (b) for random subsets of 915 protein domains from the PDUG, and the value

of these parameters $<k>$=4.61 and $<C>$=0.267 for the LUCA domains (red line). The distribution is drawn over 20,000 random selection of 915 subsets out of total 3300 PDUG domains.

Figure 1.

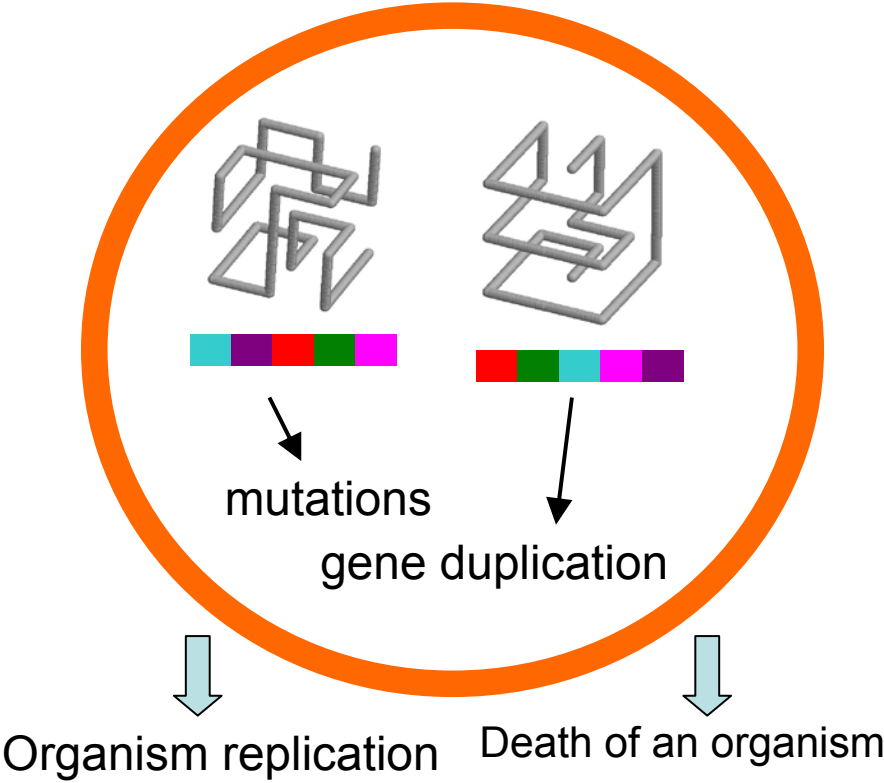

Figure 2.

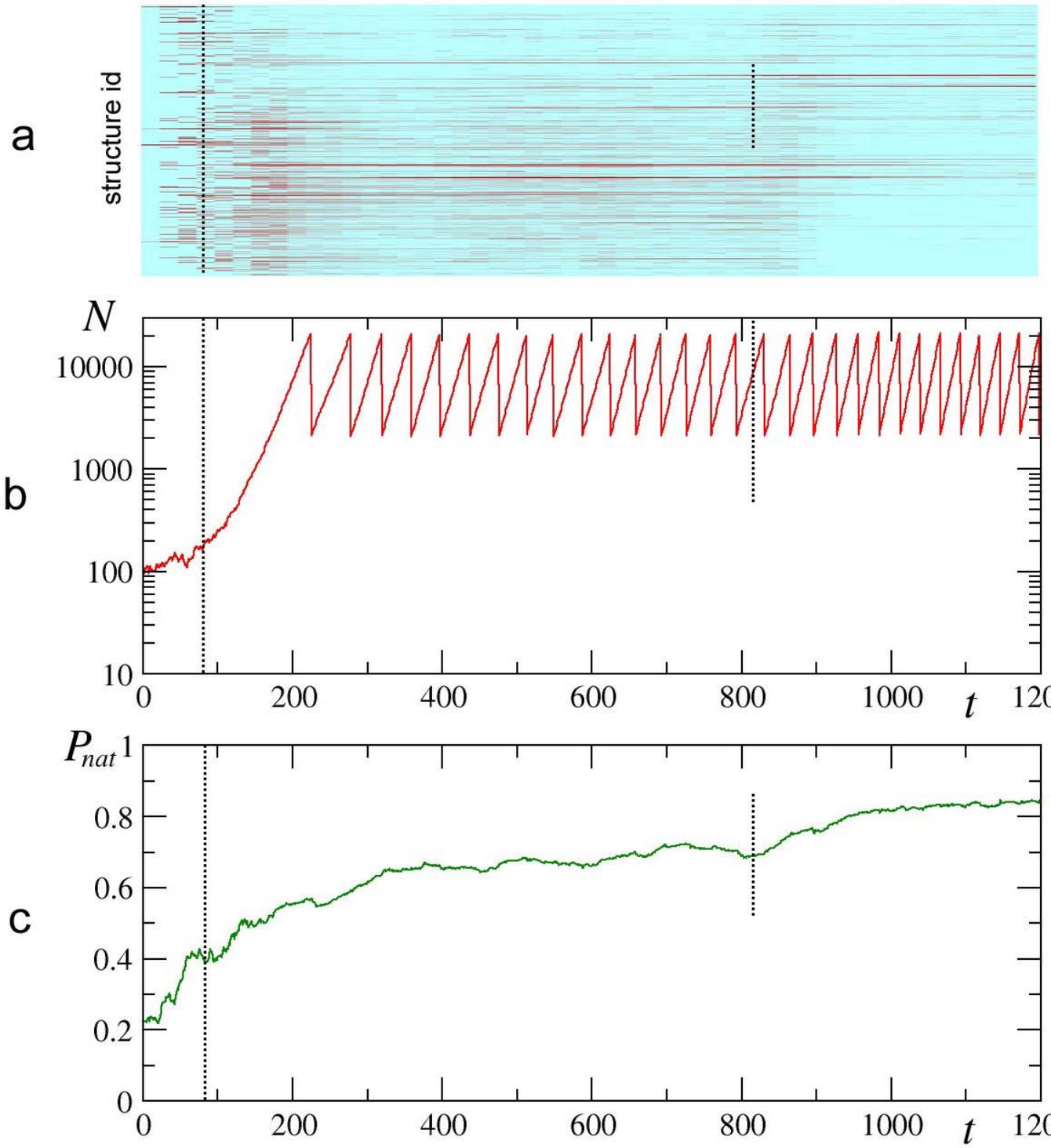

Figure 3a.

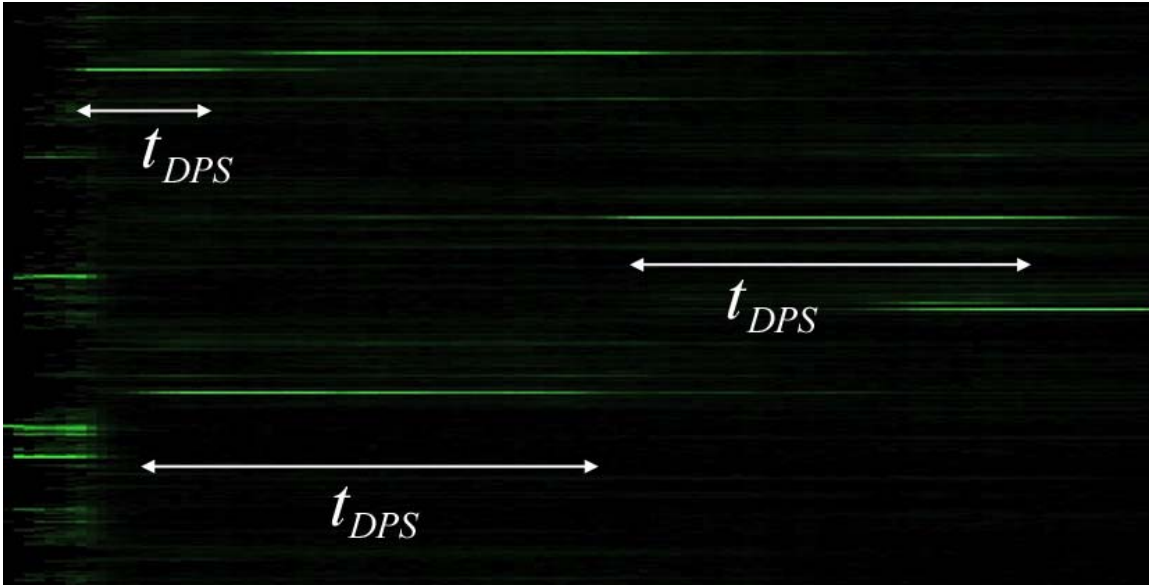

Figure 3b.

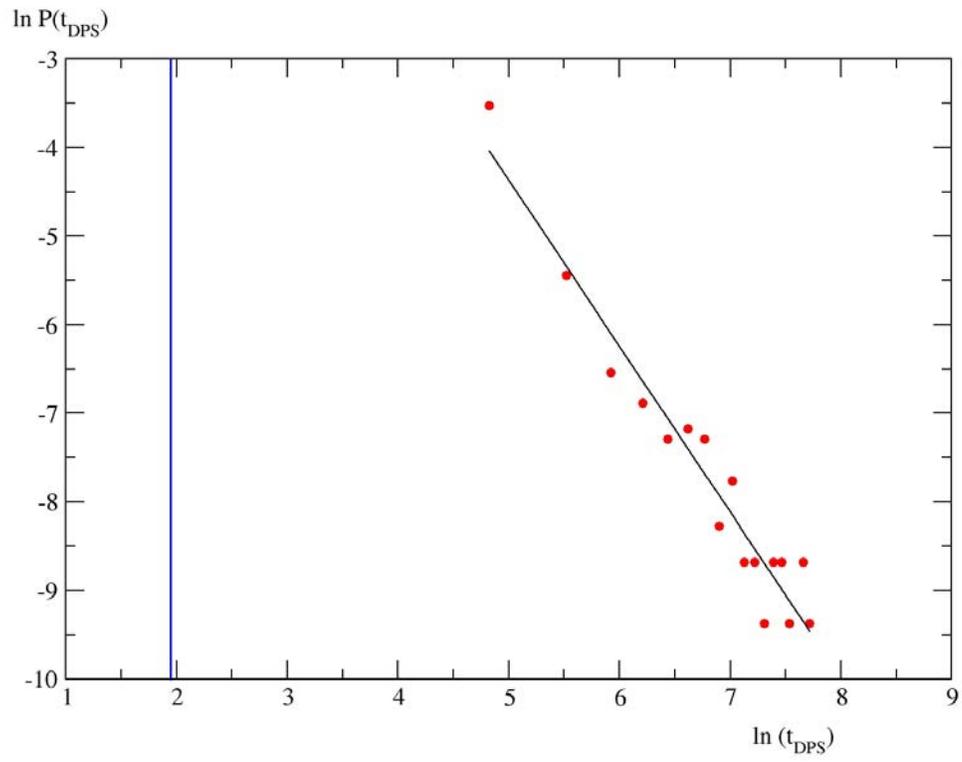

Figure 4a

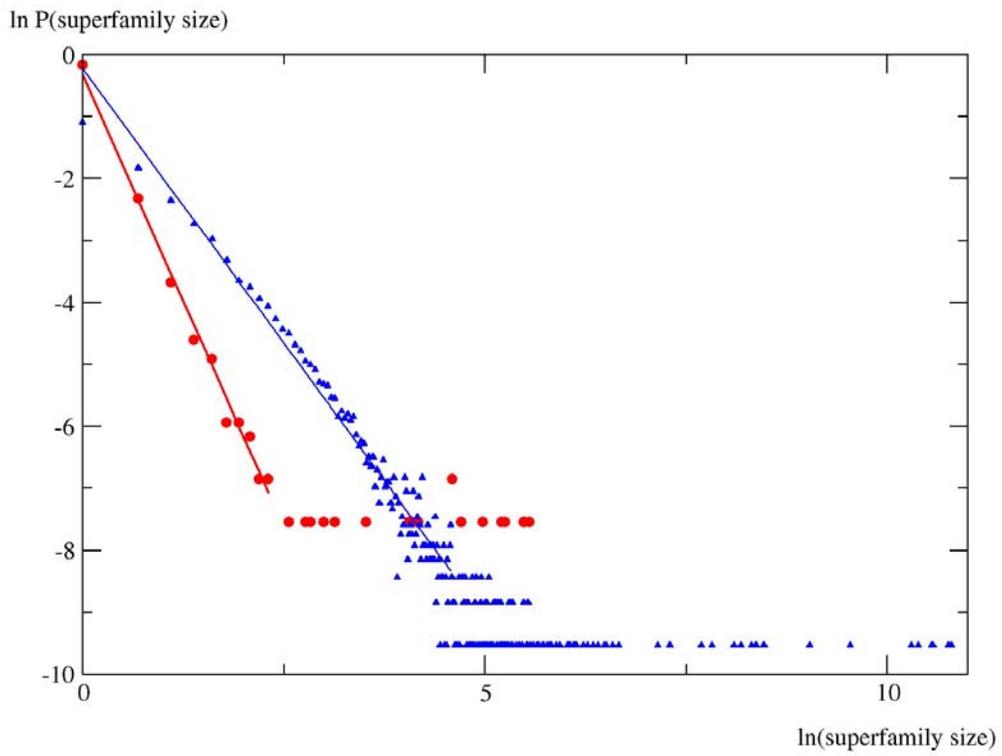

Figure 4b.

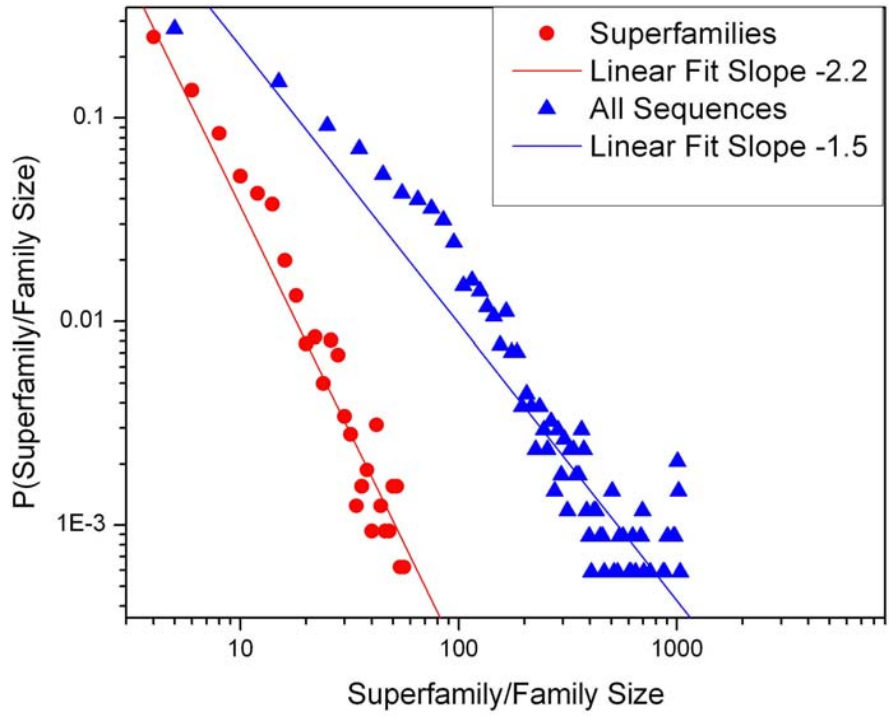

**Figure 5a.**

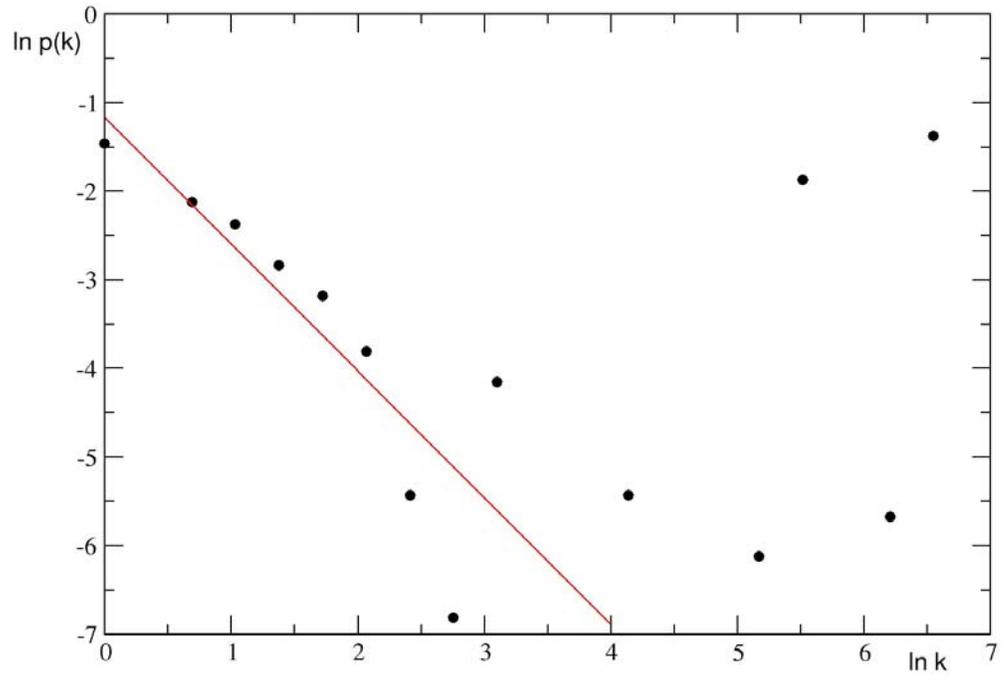

**Figure 5b.**

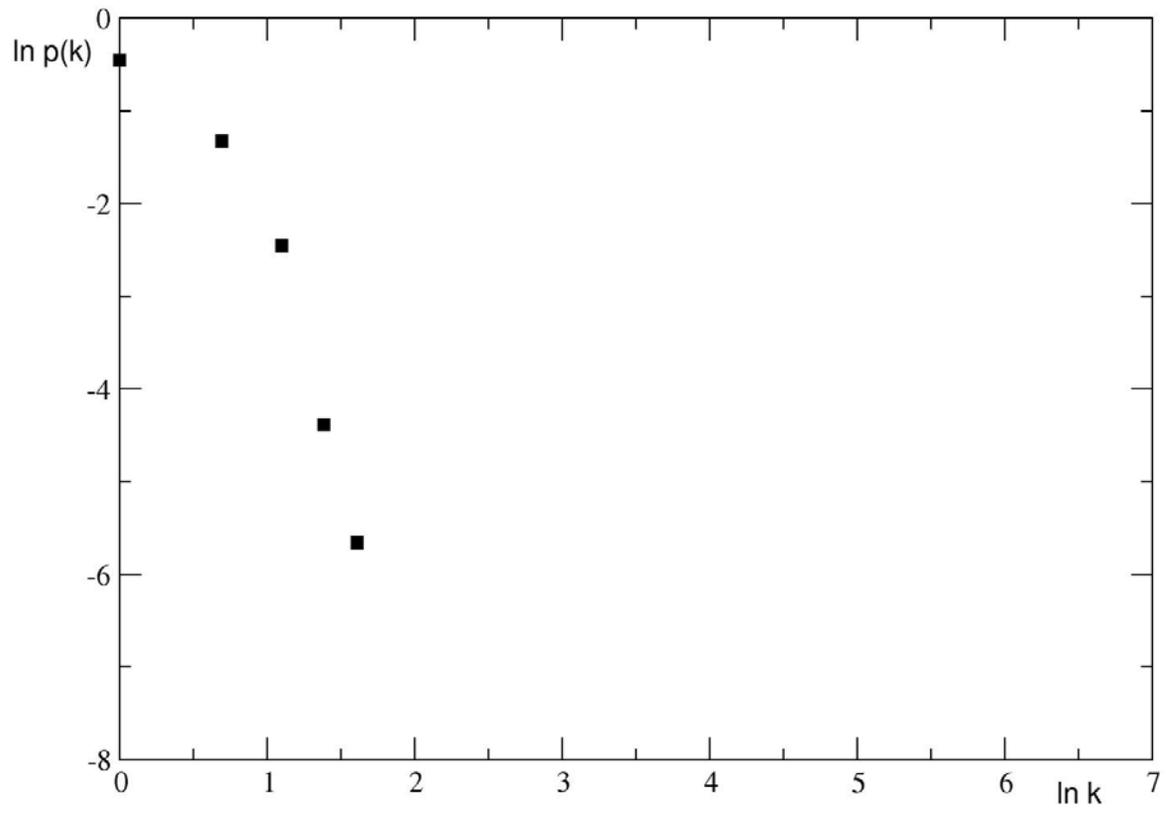

**Figure 6a**

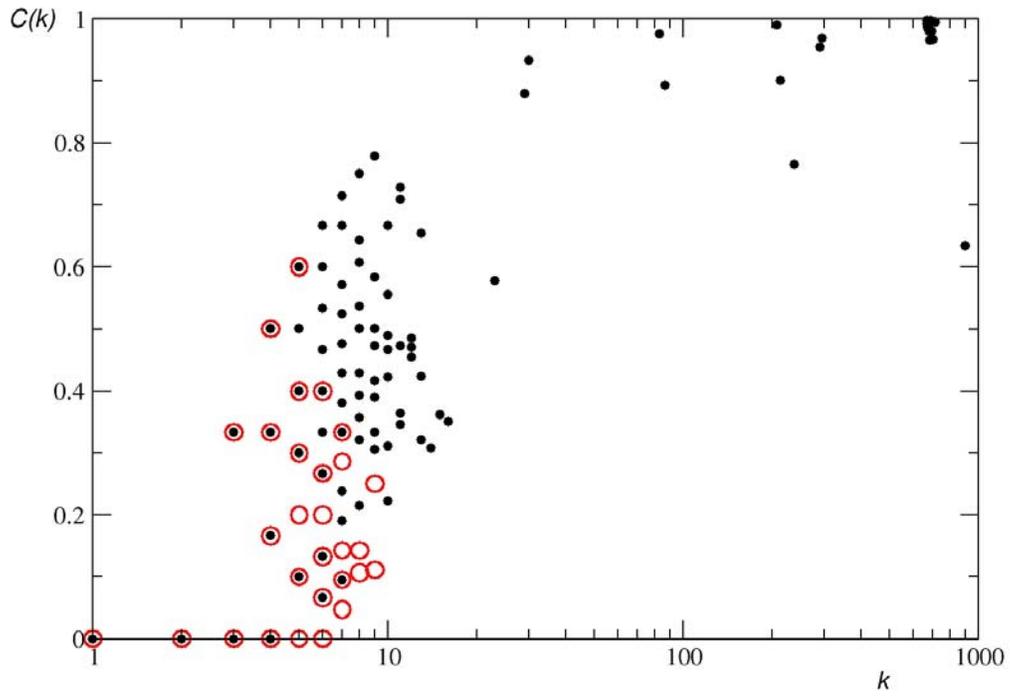

Figure 6b

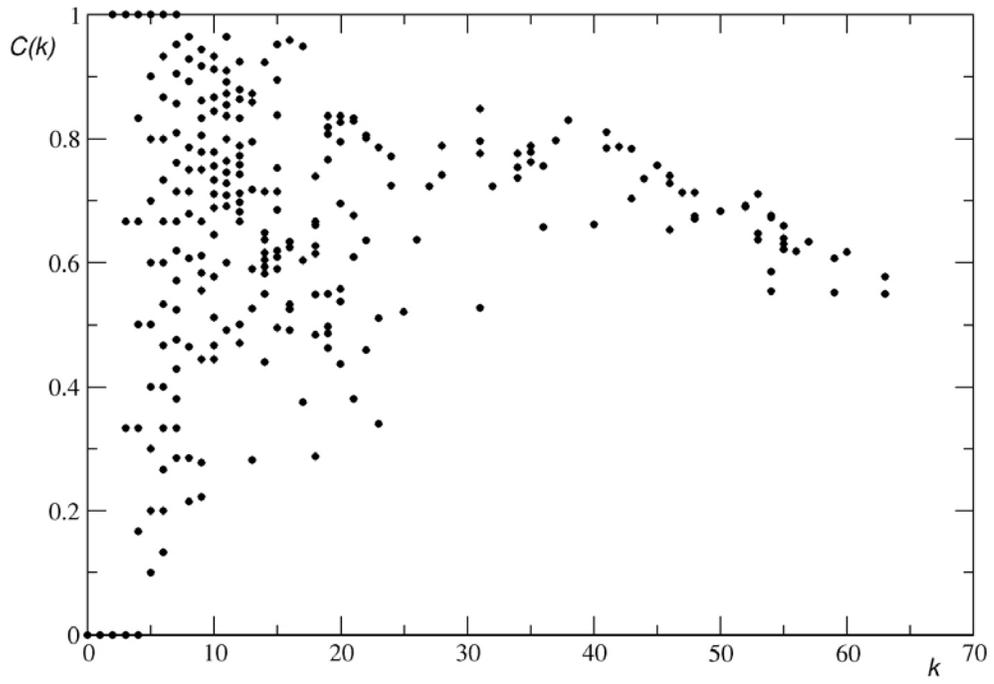

Figure 7.

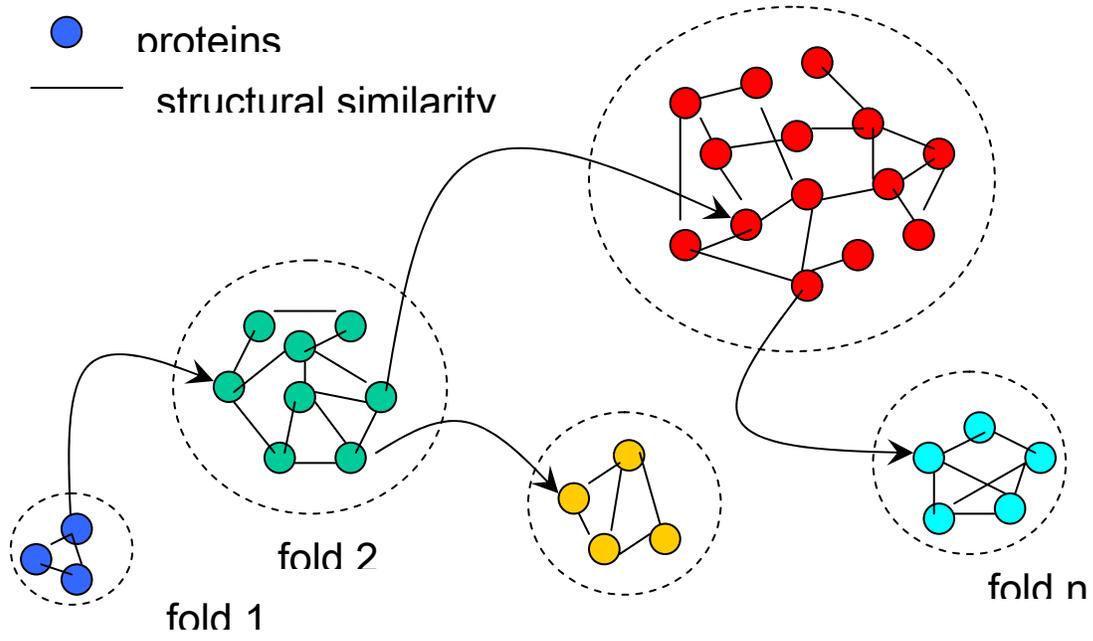

Figure 8a.

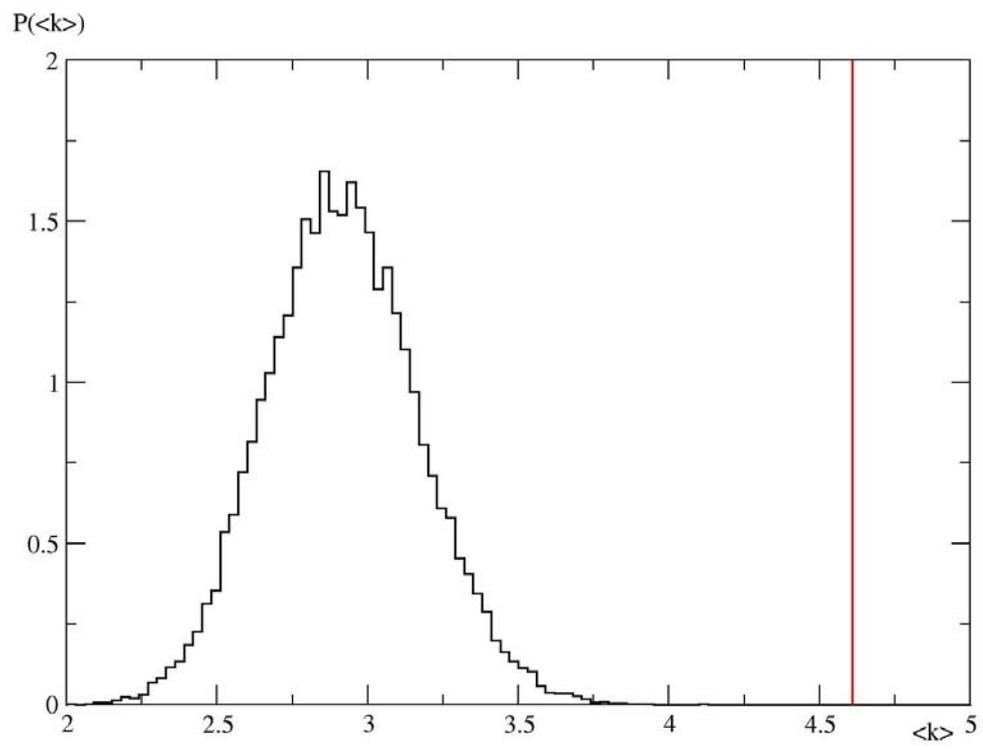

Figure 8b

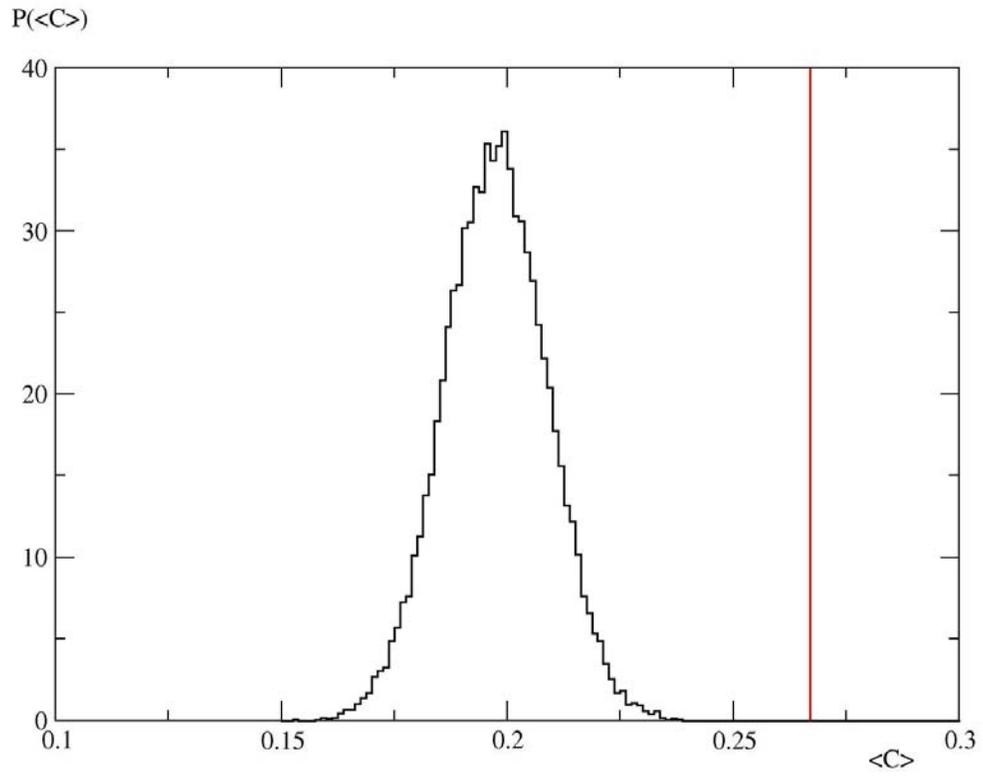